\documentclass[cits]{POS}
\usepackage{epsfig}
\usepackage{amsmath}

\newcommand{\pt}{\ensuremath{p_{T}}}
\title{Comparing energy loss phenomenology}

\ShortTitle{Comparing energy loss phenomenology}

\author{\speaker{M. van Leeuwen}\thanks{The author is indebted to the
        TECHQM collaboration for many useful discussions about the
        different energy loss schemes.}\\ 
  Utrecht University, PO Box
        80000, 3508 TA Utrecht, Netherlands\\ 
	E-mail:
        \email{m.vanleeuwen1@uu.nl}}

\abstract{High-\pt{} particle production is suppressed in heavy ion
  collisions due to parton energy loss in dense QCD matter. Here we
  present a systematic comparison of two different theoretical
  approximations to parton energy loss calculations: the opacity
  expansion and the multiple-soft scattering approximation for the
  simple case of a quark traversing a homogeneous piece of matter with
  fixed length (the TECHQM 'brick problem'), with focus on the range
  of parameters that is relevant for interpreting RHIC measurements of
  high-\pt{} hadron suppression.}

\FullConference{High-pT Physics at LHC -09\\
		 February 4 - 7 2009\\
		 Prague, Czech Republic}

\begin{document}

One of the important findings at RHIC is that high transverse momentum
(\pt) hadron production in central Au+Au collisions is suppressed
compared to (properly scaled) p+p collisions
\cite{Adcox:2001jp,Adler:2002xw}. This suppression can be attributed
to energy loss of high-\pt{} partons that traverse the hot and dense
medium formed in these collisions. An important goal of the study of
heavy ion collisions is therefore to determine the medium density from
the measured suppression by comparing the measurements to sufficiently
detailed calculations of parton energy loss and hadronisation in the
medium.

Four different theoretical descriptions of radiative energy loss in
hot and dense QCD matter are generally used. Two of the descriptions:
the opacity expansion and multiple soft scattering approach, have been
shown to be different approximations of the same basic path-length
formalism which was formulated by Baier, Dokshitzer, Muller, Peigne
and Schiff (BDMPS) \cite{Baier:1996kr} and independently by Zakharov
\cite{Zakharov:1996fv}. The higher-twist formalism was formulated by
Wang and Guo \cite{Wang:2001ifa} and starts from higher-twist terms
which are closely related to higher twist in Deeply Inelastic
Scattering calculations. It is also possible to calculate energy loss
in Hard Thermal Loop field theory, as explored by Arnold, Moore and
Yaffe (AMY) \cite{Arnold:2000dr}.

For a realistic calculation of final state hadron suppression, the
theoretical description of the fundamental energy loss process needs
to be supplemented with parton production and fragmentation and a
realistic collision geometry. Existing calculations have used
various simplified geometries, including fixed (mean) path length, 
(uniform) hard spheres and Woods-Saxon overlap
geometries. Some recent work has been done with realistic hydrodynamic
media \cite{Renk:2006pk,Renk:2006sx,Qin:2007zzf}.

For a systematic evaluation of the medium density and the
uncertainties on the density, we would like to compare the various
theoretical frameworks using the same geometry and then vary the
assumptions about the geometry in an independent way. Unfortunately,
the existing literature does not provide such a systematic
comparison. 

For example, a long-standing issue is the comparison of the opacity
expansion and the multiple-soft gluon emission approximation. The
basic energy loss kernels for both formalisms have been compared in
some detail and it was found that both formalisms could be made to
roughly agree \cite{Salgado:2003gb}, but no connection was made to the
medium density or temperature. More complete calculations of hadron suppression
using the two formalisms have been performed by two independent
groups, which makes it difficult to compare the results. For the
multiple-soft scattering approach, the PQM model implements a
Woods-Saxon geometry where the medium density is assumed to scale
with the local binary collision density \cite{Dainese:2004te}. It was
found that the average transport coefficient $\hat{q}$ that a parton
sees is 13.2 GeV$^2$/fm \cite{Adare:2008cg}, corresponding to an
estimated average medium temperature (see section \ref{sec:medium}
below) of 1 GeV. A similar calculation for the opacity expansion (GLV)
was performed by Wicks, Horowitz, Djordjevic and Gyulassy
\cite{Wicks:2005gt}, assuming the medium density to be proportional to
the local participant density. In this calculation, it is found that a
medium with about 1400 soft gluons in the transverse plane at
mid-rapidity is needed to reproduce the measured nuclear modification
factor $R_{AA}$\cite{Adare:2008cg}. This value corresponds to a
temperature of approximately 370 MeV. From this, it seems that medium
densities extracted using the two different formalisms are vastly
different (a factor 2 in $T$ corresponds to a factor 8 in density). In
the following, we will explore some of the possible origins of these
differences in detail.

\section{The TECHQM brick problem}
A general calculation of energy loss effects on high-\pt{} spectra in
heavy ion collisions involves not only the energy loss kernel which is
calculated in (perturbative) QCD, but also the input parton spectrum,
parton fragmentation, and the geometry of the reaction zone. As a
result, it can be difficult to judge whether differences in e.g. the
obtained medium density/effective temperature are generated by the
energy loss formalism itself or some of the additional assumptions.

The 'Theory Experiment Collaboration on Hot Quark Matter' (TECHQM) has
proposed a set of benchmark problems to compare different energy loss
formalisms, also known as the 'brick problem' \cite{techqm_brick}.
 The
TECHQM Brick Problem asks to calculate quark energy loss for a
fixed-length homogeneous medium (the brick). Calculations are
performed for two typical path lengths: 2 and 5 fm and 2 partons
energies 10 and 100 GeV. The following discussion will be largely
based on this problem.

\subsection{Defining a common scale: $T$}
\label{sec:medium}
In the following, we will focus on comparing the first order opacity
expansion (GLV, single-hard scattering) \cite{Gyulassy:1999zd}
approach and the multiple-soft (ASW) \cite{Salgado:2003gb}
approach. Energy loss in any scheme is expected to be governed by the
medium density $\rho$ and the path length $L$. In the two schemes
considered here, three variables are generally discussed: the average
momentum kick per radiated gluon $\langle k_{T} \rangle$, the mean
free path $\lambda$ and the path length $L$. For the multiple soft
scattering approach, the transport coefficient
$\hat{q}=\langle k_{T}^2 \rangle/\lambda$ is also often used.

In the following, the temperature $T$ will be used as a common
scale. The density of gluons in a QGP (or pure gluon gas) is
\begin{equation}
    \rho_g = \frac{16 \zeta(3)}{\pi^2}T^3 = \frac{16 \cdot 1.202}{\pi^2}T^3 = 1.94\, T^{3}.
\end{equation}
For simplicity, we approximate the medium as a pure gluon gas. The
gluon-gluon transport cross section $\sigma_{gg}$  (to logarithmic accuracy) is
\cite{Baier:2006fr}
\begin{equation}
\sigma_{gg} = \frac{9\pi\,\alpha_s^2}{2\, m_{D}^2},
\end{equation} 
where $\alpha_s$ is the strong coupling constant and $m_D$ is the Debye
screening mass ($m_{D}=\sqrt{4\pi\alpha_s}T$ for a pure gluon gas).
The mean free path is
\begin{equation} 
\lambda_g = 1/\sigma_{gg}\,\rho_g = \frac{\pi^2}{18\,\zeta(3)\,\alpha_s}\frac{1}{T}.
\end{equation} 
We further take $\langle k_T^2 \rangle = m_{D}^2$. This ties all the parameters in the
calculations to $T$.


\section{Comparing energy loss -- importance of distribution width}
\begin{figure}
  \epsfig{file=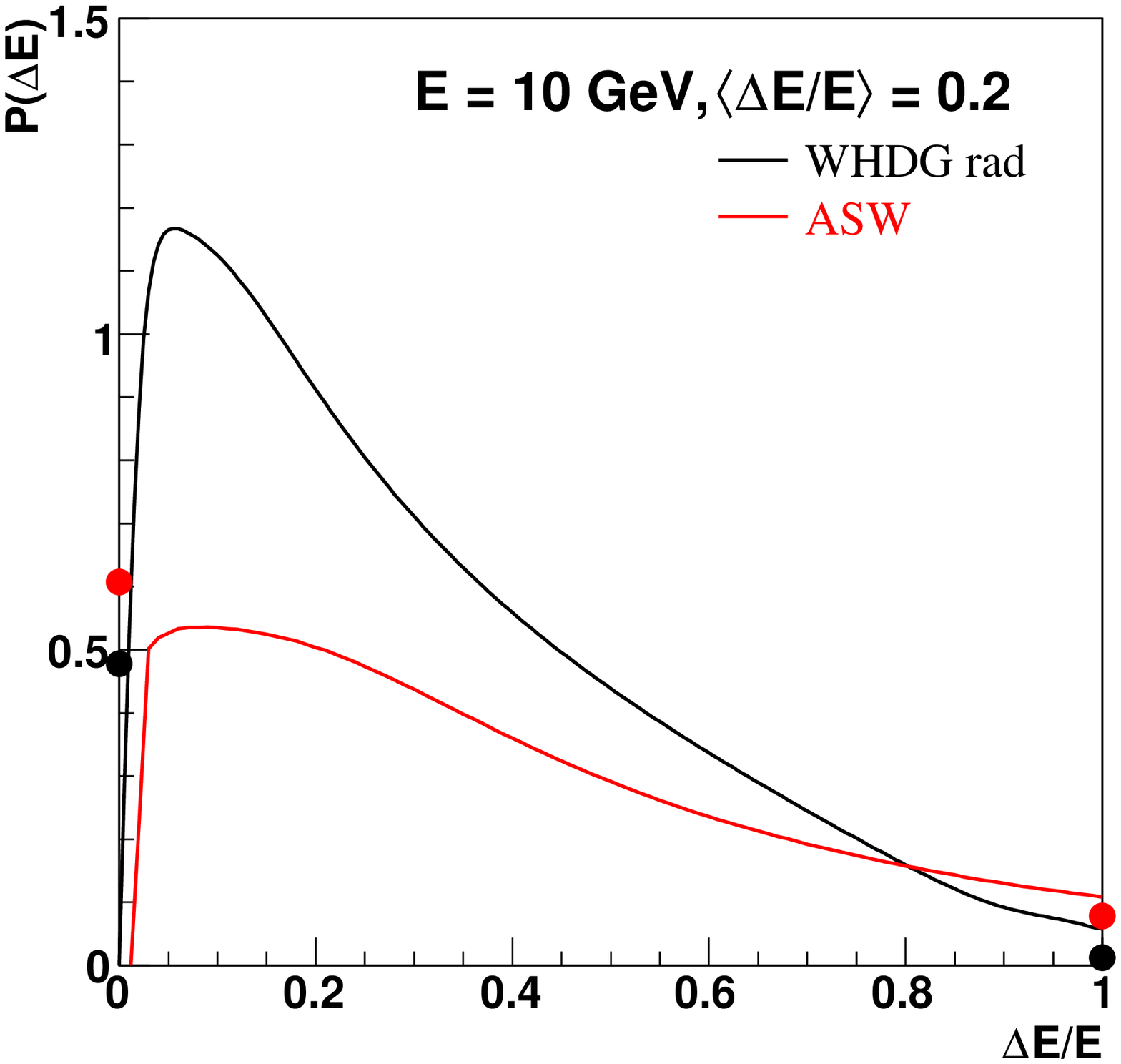,width=0.5\textwidth}
  \epsfig{file=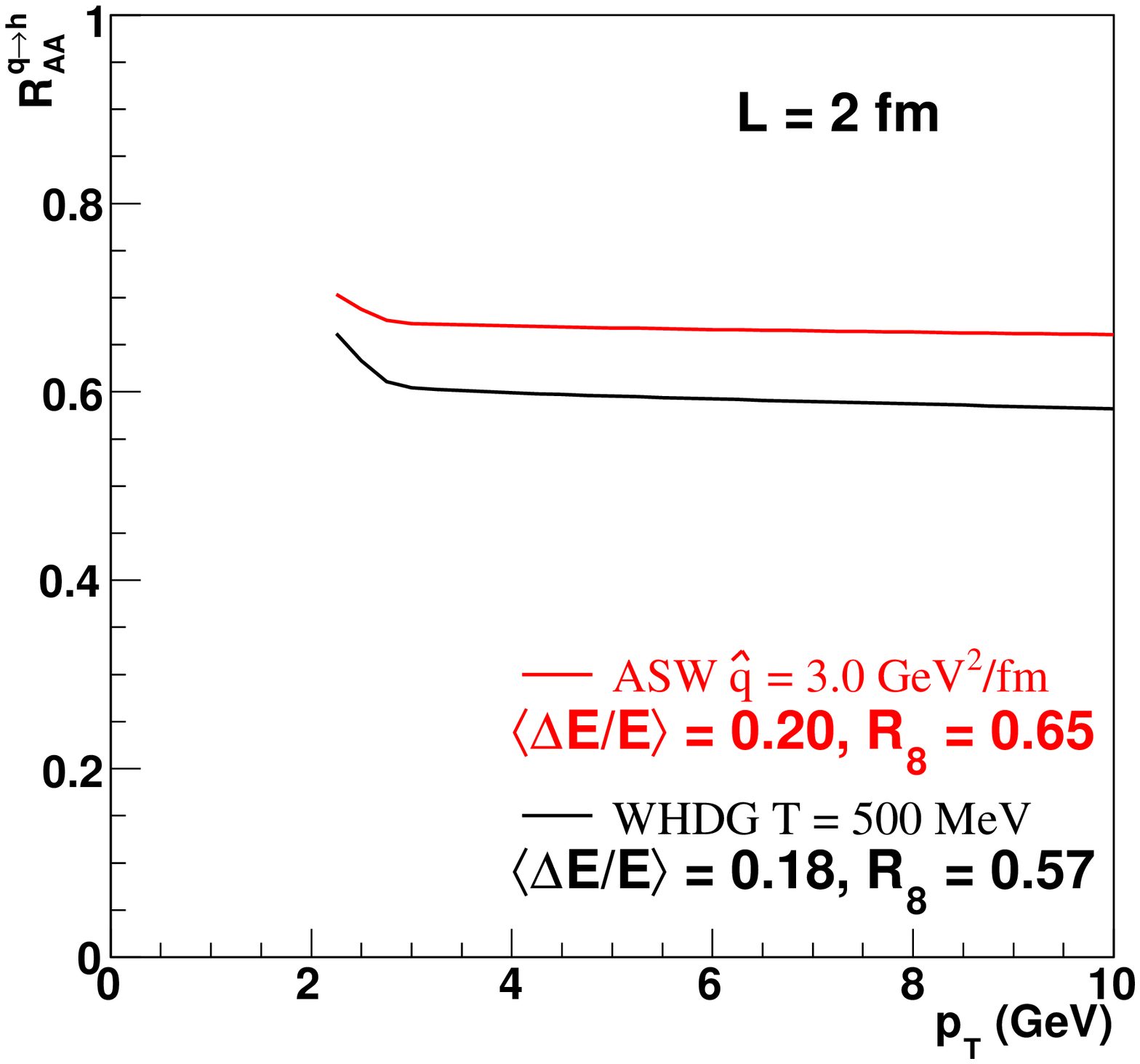,width=0.5\textwidth}
  \caption{\label{fig:RAA_de02}Left panel: energy loss distribution
  for a quark with $E$ = 10 GeV in a homogeneous medium of $L=2$ fm
  and $\langle \Delta E/E \rangle \simeq 0.2$, from the multiple-soft
  scattering ASW approach and the WHDG opacity expansion. The
  points on the left and right axis give the discrete probabilities to
  lose no energy and to lose the entire energy. Right
  panel: Nuclear modification factor for pions from quark
  fragmentation, using the energy loss distributions in the left panel
  and KKP fragmentation.}
\end{figure}
Figure \ref{fig:RAA_de02}, left panel, compares the probability
(density) distribution for energy loss $P(\Delta E)$ for the WHDG opacity
expansion (black curve) \cite{Wicks:2005gt} and ASW multiple soft scattering
approach (red curve) \cite{Salgado:2003gb} for a 10 GeV quark and
$L=2$ fm. For this comparison, the medium densities have been set to
obtain an average fractional energy loss of 0.2 for both models ($T =
500$ MeV and $\hat{q} = 3$ GeV$^2$/fm). The two different approaches
give significantly different energy loss distributions, starting from
the probability to lose no energy at all $P(0)$, which is ~0.5 for the
opacity expansion and ~0.6 for the multiple soft scattering approximation. The
continuous part the distribution shows correspondingly larger
probabilities for WHDG than ASW. In addition, the opacity expansion
curve is more strongly peaked towards small energy losses.

A mean fractional energy loss of 0.2 was chosen for
Fig.~\ref{fig:RAA_de02}, because for a power law spectrum with $n=8$,
this is expected to generate $R_{AA} \sim (1-0.2)^{n-1} = 0.21$ close
to the measured value of 0.2--0.25 at
RHIC~\cite{Adcox:2001jp,Adler:2002xw,Adare:2008cg}. The right panel
of Fig.~\ref{fig:RAA_de02} shows $R_{AA}$ calculated for quark fragments,
using a realistic (LO pQCD) parton spectrum and KKP fragmentation
\cite{Kniehl:2000hk}. The observed values of $R_{AA}^q$ is much larger
than the naively expected value of 0.21, due to the broad distribution
$P(\Delta E)$. In addition, it is seen that $R_{AA}$ is smaller for
the opacity expansion formalism than in the multiple-soft scattering approximation, as
one would expect from the smaller value of $P(0)$ ($R_{AA}$ has to be
greater than or equal to $P(0)$).

A better estimate of $R_{AA}$ is the weighted average:
\begin{equation}
R_n = \int_0^1 d\epsilon (1-\epsilon)^n P(\epsilon),
\end{equation}
where $\epsilon = \Delta E/E$. The calculated values of $R_8$ are shown
in the figure and are in approximate agreement with $R_{AA}^q$.

\section{Single-hard and multiple-soft scattering}

\begin{figure}
\begin{center}
  \epsfig{file=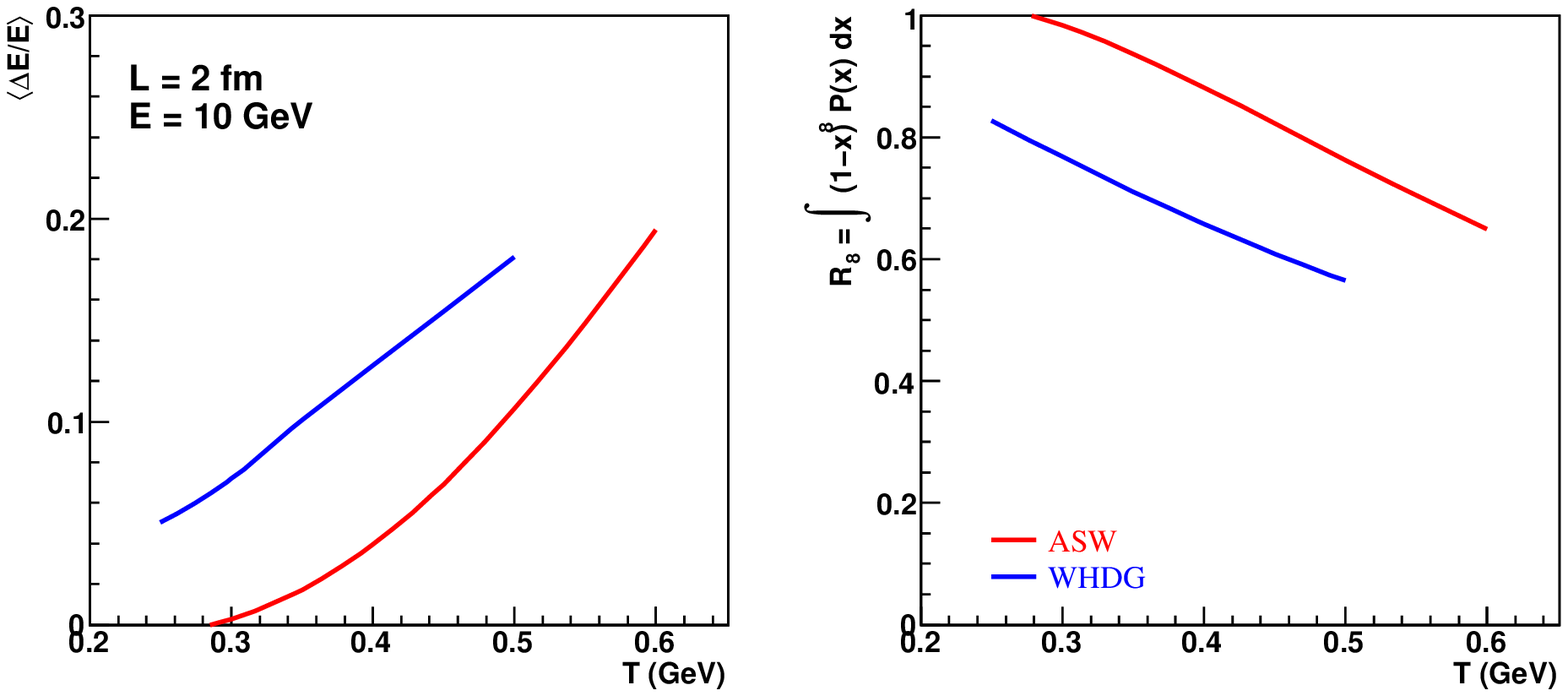,width=0.8\textwidth}
  \epsfig{file=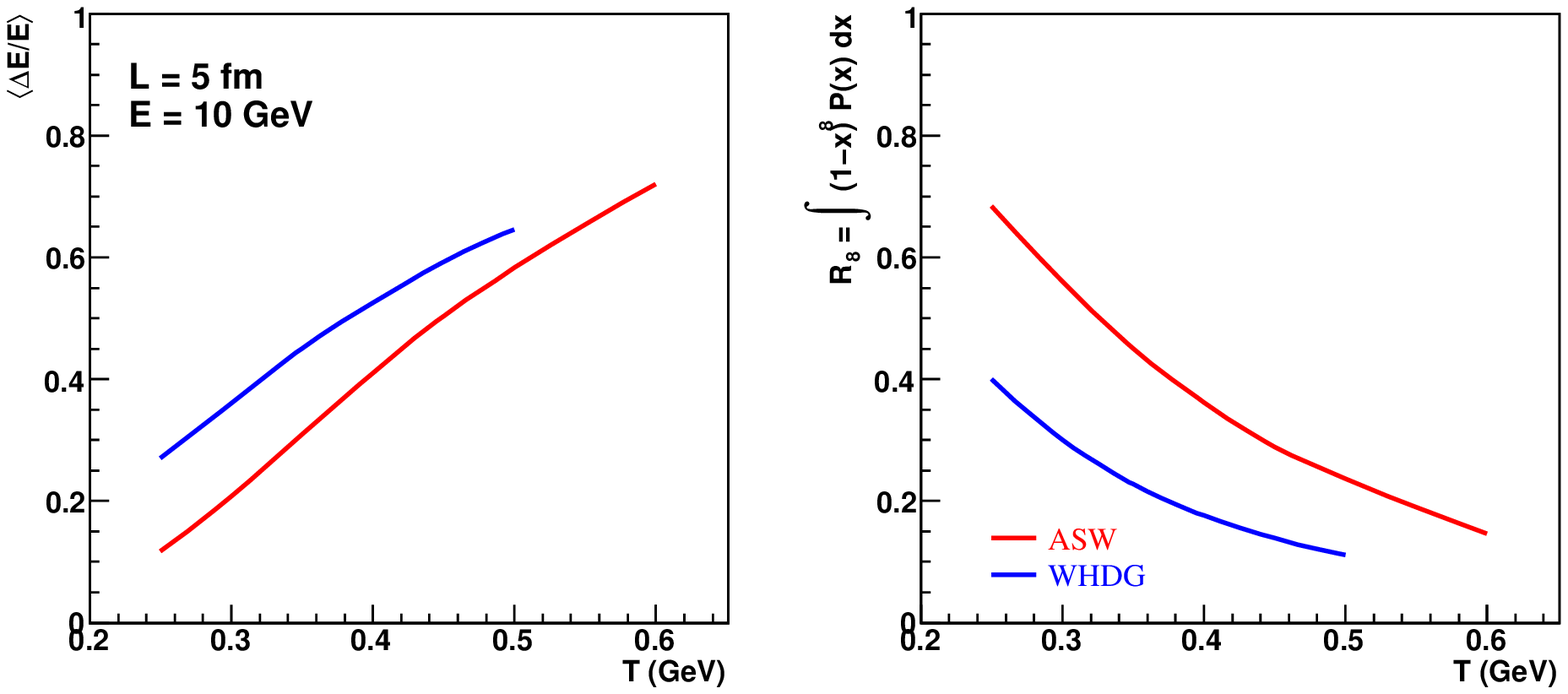,width=0.8\textwidth}
\end{center}
  \caption{\label{fig:dEvsT}Mean fractional energy loss (left panels)
  and $R_8$ (right panel) as a function of medium temperature for a
  quark passing through a homogeneous medium of length $L = 2$ fm
  (upper panels) and $L=5$ fm (lower panels). The red curves shows the
  result in the multiple-soft scattering approach
  (ASW)\cite{Salgado:2003gb} and the blue curves are for the
  opacity expansion (WHDG)\cite{Wicks:2005gt}.}
\end{figure}
Figure \ref{fig:dEvsT} shows a comparison of energy loss in the
opacity expansion and the multiple soft scattering formalism as a
function of the temperature of the medium. The upper panels show
results for $L=2$ fm and the lower panels for $L=5$ fm. The values of
$\hat{q}$ for the ASW calculation and $\mu, \lambda$ for the WHDG
calculation were calculated using the equations in Section
\ref{sec:medium}. It can be clearly seen in the figure that the WHDG
formalism typically leads to a larger mean energy loss or smaller
$R_{AA}$ for a given temperature. This might be due to the larger
impact of the high-energy tail in the energy spectrum of radiated
gluons in the opacity expansion than in the multiple soft scattering
approximation \cite{Salgado:2003gb}. The differences are more
pronounced at low temperatures and short path lengths than at high
temperatures and long path lengths. For $L=2$ fm, the fractional
energy loss reaches 0.2 at $T=500$ MeV and $T=600$ MeV for WHDG and
ASW respectively. The corresponding values of $R_8$ are close to
0.6. With a medium length $L=5$, the energy loss is much larger,
reaching $R_8 \sim 0.2$ at $T=375$ MeV for WHDG and $T=530$ MeV
($\hat{q}=$ 1.9 GeV$^2$/fm) for ASW. Clearly, large path lengths are
needed to reach the value $R_{AA} \sim 0.2$ that is observed at RHIC. The
corresponding mean energy loss is also large: $\epsilon \sim 0.5$.

It is interesting to note that using the brick geometry, the
difference between the temperature needed to reach $R_{AA}=0.2$ in the
GLV and BDMPS formalism is about a factor 1.5, while some earlier
studies indicated factors of 2--2.5 (see introduction).

\section{Note on geometry}
The space-time geometry of the collision is an important ingredient of
realistic calculations of hadron suppression due to parton energy
loss. In the above, a fixed path length and homogeneous medium were
used to set up a simple benchmark comparison. Most existing
calculations in the literature use somewhat more realistic, parametrised
geometries which start from hard spheres or the Woods-Saxon density
profile of cold nuclei. Only very recent calculations
use realistic geometries from hydrodynamic calculations
\cite{Renk:2006pk,Renk:2006sx,Qin:2007zzf}.

\begin{figure}
\epsfig{file=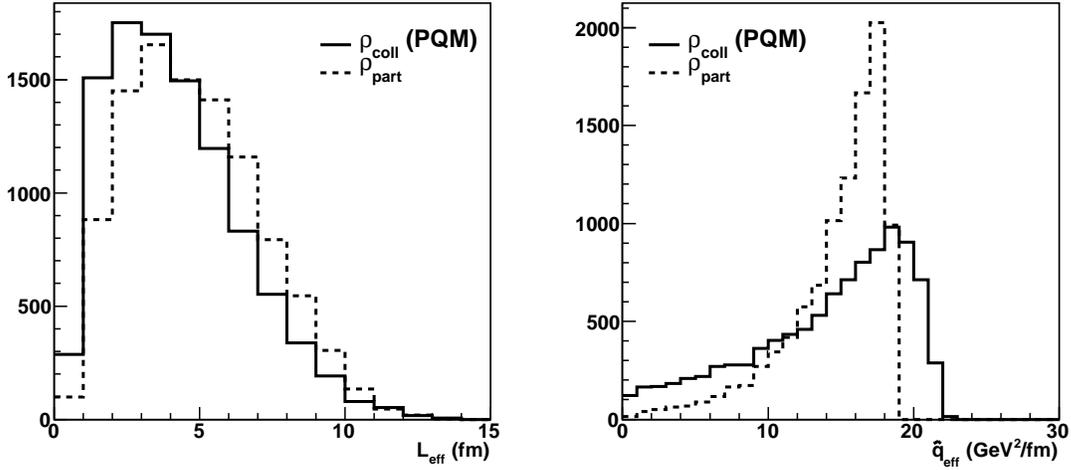,width=\textwidth}
\caption{\label{fig:L_q_comp}Distributions of effective path length
  $L_{eff}$ and transport coefficient $\hat{q}_{eff}$ in the case
  where the local medium density is proportional to the density of
  hard scatterings $\rho_{coll}$ (solid line) and the density of
  participant nucleons $\rho_{part}$ (dashed line) for 0-10\% central
  Au+Au collisions. The proportionality constants for the medium
  density scaling are given in the text.}
\end{figure}
In particular, the PQM model \cite{Dainese:2004te} uses quenching
weights from the multiple-soft scattering approximation with a density
profile that is proportional to the local density of hard scatterings
$\rho_{coll}$, while the WHDG calculations \cite{Wicks:2005gt} use the
opacity expansion with a density profile that is proportional to the
local density of participants $\rho_{part}$. Figure \ref{fig:L_q_comp}
compares the distributions of effective path lengths $L_{eff}$ and
transport coefficient $\hat{q}$, for a medium medium density scaling
with $\rho_{coll}$ and $\rho_{part}$. $L_{eff}$ and $\hat{q}_{eff}$
are calculated using the same averaging procedure as in the PQM model:
\begin{equation}
L_{eff} = \frac{2\,I_1}{I_0}, \; \hat{q}_{eff} = \frac{{I_0}^2}{2\,I_1},
\end{equation}
with 
\begin{equation}
I_{n} = \int_0^{\infty} du\, u^n\, \hat{q}(u),
\end{equation}
where $\hat{q}(u)$ is the local density and the integral runs along the
path of the parton through the medium. The points of origin for the
partons follow a $\rho_{coll}$ distribution in both cases.

The density distributions as a function of the position the transverse
plane $\vec{s}$ and the impact parameter $\vec{b}$ are calculated from
the thickness functions $T_{A}$ using
\begin{equation}
\rho_{coll}(\vec{s};\vec{b}) = T_{A}(\vec{s}-\tfrac{1}{2}\vec{b})\, T_{B}(\vec{s}+\tfrac{1}{2}\vec{b}),
\end{equation}

\begin{equation}
\rho_{part}(\vec{s};\vec{b}) = T_{A}(\vec{s}-\tfrac{1}{2}\vec{b})\, 
\left(1 - e^{-\sigma\, T_B(\vec{s}+\tfrac{1}{2}\vec{b})} \right) + 
T_{B}(\vec{s}+\tfrac{1}{2}\vec{b})\,
\left(1 - e^{-\sigma\, T_A(\vec{s}-\tfrac{1}{2}\vec{b})} \right).
\end{equation}

The terms in brackets in the equation for $\rho_{part}$ are needed to
cut off the naive $\rho_{part} \propto T_A + T_B$ in cases where the
density in one of the nuclei is close to zero. These terms are the
Poisson probability to have at least one struck nucleon in the 'other'
nucleus ($\sigma$ is the total inelastic cross section for a
proton-proton collision, 42 mb at RHIC energies).

There is one free scaling parameter in both distributions. For the
case of $\rho_{coll}$ scaling, the parameter has been set to the same
value as in the PQM model ($k_{coll} = 5 \cdot 10^6\;\mathrm{fm}/A^2$
\cite{Dainese:2004te}). The value for $\rho_{part}$ was chosen such that the average
$\hat{q}_{eff}$ was similar to the $\rho_{coll}$ case ($k_{part} = 2.3
\cdot 10^4\;\mathrm{fm^{-1}}/A$).

The qualitative difference between the  $\rho_{part}$ geometry and the $\rho_{coll}$ geometry is that the participant density is more
uniform than the collision density. This leads to a larger $L_{eff}$
and a more sharply peaked distribution for $\hat{q}_{eff}$ than in the
case where the density scales with $\rho_{coll}$. The corresponding
suppression is therefore expected to be larger for a participant
density geometry than for the collision density scaling. This may
partially explain the differences between the medium density as
determined from the WHDG model and from the PQM model.

\section{Discussion and outlook}
A full calculation of hadron suppression due to parton energy loss in
heavy ion collisons contains a number of ingredients. Existing
calculations have found a large range of different values for the
medium density that is needed to describe the measured suppression at
RHIC. 

When comparing radiative energy loss for a fixed pathlength in a
homogeneous medium in the first order approximation of the opacity
expansion and the multiple soft scattering approximation, the
differences are found to be sizable, with a significant dependence on
path length and medium density or temperature. For example, $R_{AA}
\sim 0.2$ is reached for a quark in a gluon gas of temperature $T=375$
MeV in the opacity expansion and $T=530$ MeV for the multiple soft
scattering approach. In both calculations, the average energy loss is
large (approx. 50\%) and the probability distribution is very
broad. This is an important to keep in mind for phenomenolgy, because
it means that one cannot think in terms of 'typical' or 'average'
energy loss. In addition, the formalisms to calculate energy loss use
approximations that are applicable for small fractional energy
loss. It would be interesting to estimate the uncertainties from these
assumptions.

Another important aspect of energy loss models is the medium density
profile. Two different assumptions are commonly found in the
literature: the local medium density is either proportional to the
local participant density $\rho_{part}$ or to the local collision
density $\rho_{coll}$. The latter approach leads to shorter effective
path lengths and larger fraction of partons that experience a small
effective medium density. Both effects lead to a smaller energy loss
for a $\rho_{coll}$ density profile than for a $\rho_{part}$ profile with
the same central density.

Future work will concentrate on using the different energy loss
formalisms in a more realistic geometry. We also aim to extend the
comparison to the Higher Twist and Hard Thermal Loop
formalisms.

\bibliographystyle{epj_pos}
\bibliography{eloss_models_mvanleeuwen}
\end{document}